\documentclass[twocolumn,preprintnumbers,amsmath,amssymb,prb]{revtex4}

\usepackage{graphicx}
\usepackage{dcolumn}
\usepackage{bm}
\usepackage{color}

\newcommand{\highlight}{\color{black}}

\begin{document}

\title{Enhanced circular dichroism {\highlight via slow-light in dispersive structured media}}

\author{Jesper Pedersen and Niels Asger Mortensen}
\affiliation{MIC -- Department of Micro and Nanotechnology,
             NanoDTU, Technical University of Denmark, Building 345east,
             DK-2800 Kongens Lyngby, Denmark}

\date{\today}
\begin{abstract}
Circular dichroism (CD) is in widespread use as a means of
determining enantiomeric excess. We show how slow-light phenomena
in {\highlight dispersive structured media} allow for a reduction in the required optical
path length of an order of magnitude. The same ideas may be used
to enhance the sensitivity of CD measurements while maintaining
the same optical path length through the sample.
Finally, the sensitivity may be enhanced in frequency regimes
where CD data is typically not accessible due to a modest chiral
response of the enantiomers.
\end{abstract}

\maketitle

The determination of enantiomeric excess is of vital importance in
e.g. the pharmaceutical and biochemical industries. While most
biological chiral molecules only occur naturally with a single
handedness, industrial production of enantiomerically pure
products is an extremely difficult  and challenging
task. While one enantiomer may have the desired effect, the effect
of the other enantiomer may be less pronounced or may even be
toxic in some cases, placing a strong requirement on the
development of tools for determination and
quantification of enantiomeric excess. One widespread and
commercially available tool for measuring enantiomeric excess is
the Ultraviolet-Circular Dichroism (UV-CD) method, which combines
ordinary UV spectroscopy with a measurement of circular dichroism
(CD) to determine the enantiomeric excess of a sample.
\cite{Chen2003}
{\highlight UV light is usually required, because most relevant
biochemical molecules of interest absorb predominantly in the UV region.}
While UV spectroscopy provides information on the
total concentration $c_R+c_S$ of both types of enantiomers, CD
measurements provide information on the difference $c_R-c_S$. A
combined UV-CD measurement thus immediately gives the enantiomeric
excess $(c_R-c_S)/(c_R+c_S)$ of a sample. With the on-going
emphasis on miniaturization of chemical analysis systems
\cite{Janasek2006} there is an increased emphasis on
optofluidics\cite{Psaltis:2006,Monat:2007a,Erickson2007} and the
integration of optics with lab-on-a-chip
microsystems.\cite{Verpoorte2003} In particular, there is a
strong desire for enhancement of the sensitivity of optical
measurements, in order to accommodate the strongly reduced optical
path length in such systems.\cite{Mortensen2007a} In
this Letter we theoretically show how the introduction of
{\highlight dispersive structured media} in CD measurements may allow for a reduction
in the required optical path length of an order of magnitude,
while maintaining the same signal-to-noise ratio.

\begin{figure}[b!]
\begin{center}
\includegraphics[width=\columnwidth, trim = 0 0 0 0, clip]
{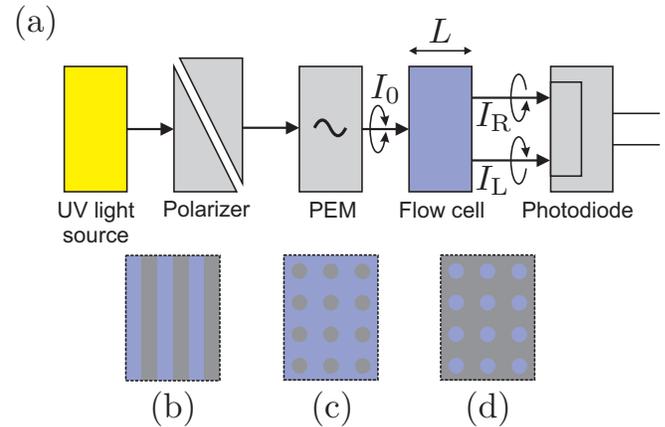}
\end{center}
 \caption{(Color online) (a) Schematics of a measurement of circular
 dichroism, in which the difference in absorbance of left and
 right circularly polarized light is measured. If the sample contains
 a chiral liquid, $I^{(L)}\neq I^{(R)}$. (b)-(d) Examples of photonic
 crystals facilitating slow-light phenomena, strongly enhancing the
 light-matter interactions in the sample.}
\label{fig:setup}
\end{figure}
Consider a typical setup for measuring chiral dichroism, as
illustrated schematically in Fig.~\ref{fig:setup}a.\cite{UV-CD}
Here, light from a UV source is directed to a polarizer and then
modulated between left- and right-handed circularly polarized
light by a photoelastic modulator (PEM). The resulting circularly
polarized light of intensity $I_0$ is then passed through a flow
cell of length $L$, and the difference in transmitted intensity,
$I_\mathrm{L/R}$ of each polarization is measured. The absorption
parameter of the chiral liquid is given as $\alpha_i$ for left
($i=L$) and right ($i=R$) circularly polarized light,
respectively. To lowest order in the
concentrations $\alpha_i\propto c_i$ with the constants of
proportionality known as the extinction coefficients. We then
find, using Beer's law, that the difference in intensity of the
transmitted light, $\Delta I\equiv I_\mathrm{L}-I_\mathrm{R}$, is
given as
\begin{equation}
\Delta I =
I_0\left(e^{-\alpha_\mathrm{L}L}-e^{-\alpha_\mathrm{R}L} \right).
\end{equation}
To maximize the signal to noise ratio for such a measurement, $L$ should
be chosen so that
\begin{equation}
L = \frac{\ln\left(\alpha_\mathrm{L}/\alpha_\mathrm{R}\right)}{\alpha_\mathrm{L}-\alpha_\mathrm{R}},
\end{equation}
for which $\Delta I/I_0$ is at an extremum. In typical,
commercially available UV-CD apparatus, $L\approx
1$~$\sim25$~mm, \cite{JASCO} {\highlight though this may not
necessarily be the optimum length.}

Consider now the situation in which the liquid is embedded in a
{\highlight dispersive} photonic crystal, as illustrated schematically in
Figs.~\ref{fig:setup}b-d. Due to slow-light phenomena the
light-matter interactions are enhanced in the photonic crystal, as
shown recently by Mortensen and Xiao in
Refs.~\onlinecite{Mortensen2007} and \onlinecite{Mortensen2007a}.
This enhancement may be quantified via an enhancement factor
$\gamma\equiv\alpha_e/\alpha$, where $\alpha_e$ is the absorption
parameter when the photonic crystal is present. The dimensionless
enhancement factor is given
as\cite{Mortensen2007,Mortensen2007a}
\begin{equation}\label{eq:gamma}
\gamma = f\times\frac{c/n_l}{v_g},
\end{equation}
where $n_l$ is the refractive index of the liquid, $v_g$ is the
group velocity and $f$ is the filling factor giving the relative
optical overlap with the liquid. In the absence of the photonic
crystal, $v_g=c/n_l$ and consequently $\gamma=1$.
Eq.~(\ref{eq:gamma}) has been derived in the case of a photonic
crystal infiltrated by a non-chiral liquid with absorption
coefficient $\alpha$.\cite{Mortensen2007} To justify its use also
for weakly chiral liquids we recall the constitutive relations for
the chiral problem
\begin{subequations}
\begin{eqnarray}
{\mathbf D} = \epsilon({\mathbf r}) {\mathbf E}({\mathbf r}) + i\xi({\mathbf r}) {\mathbf B}({\mathbf r}),\\
{\mathbf H} = \frac{1}{\mu({\mathbf r})}{\mathbf B }
({\mathbf r}) + i\xi({\mathbf r}) {\mathbf
E}({\mathbf r}),
\end{eqnarray}
\end{subequations}
where $\xi$ is a parameter quantifying the chiral strength of the
liquid. Combining Maxwell's equations we arrive at the standard
wave equation for the non-chiral problem to first order in $\xi$.
In other words, standard electromagnetic perturbation theory shows
that to lowest order in $\xi$ no frequency shifts are introduced
by chirality. Consequently, we may in a first order approximation
use the same group velocity $v_g$ for both directions of circular
polarization of the incident light. Including slow-light phenomena
in the previous discussion via the enhancement factor, we thus
find that the maximum signal to noise ratio is found when
\begin{equation}\label{eq:gammaL}
\gamma L =\frac{\ln\left(\alpha_\mathrm{L}/\alpha_\mathrm{R}\right)}{\alpha_\mathrm{L}-\alpha_\mathrm{R}}.
\end{equation}
This result suggests, quite intuitively, that a large
enhancement factor facilitates miniaturization of circular dichroism
setups. Such a miniaturization is crucial for possible future implementations of
chirality measurements in lab-on-a-chip systems.
{\highlight If one considers lower concentrations of chiral molecules, then the optimum
value of $\gamma L$ will increase, and, conversely, a larger value of $\gamma L$
will increase the measurement sensitivity at low concentrations. Thus,} the
sensitivity of existing CD measurements may be increased, while
maintaining the same optical path length through the sample. Also, it may allow
for tuning the setup to accommodate different types of molecules,
by dynamically altering the dispersion relation. This
may be accomplished, e.q., by tuning the wavelength of the incident
light or the refractive index of the materials of the {\highlight structure}.

\begin{figure}
\begin{center}
\includegraphics[width=.95\linewidth, trim = 0 0 0 0, clip]
{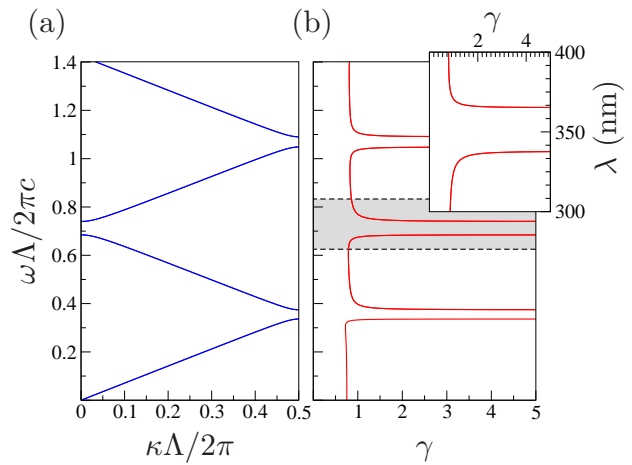}
\end{center}
 \caption{(Color online) Bragg stack of period $\Lambda=a_l+a_0$,
 with $n_l=1.33$, $n_0=1.7$, $a_l=0.8\Lambda$, and $a_0=0.2\Lambda$.
 (a) Photonic band structure for normal incidence of either TM or
 TE polarized light. (b) Corresponding enhancement factor $\gamma$,
 which exhibits marked peaks near the photonic band-gap edges,
 where it increases several orders of magnitude. Inset: Enhancement
 factor as a function of wavelength in the UV region, for a structure
 with $\Lambda=250$~nm. The wavelength range corresponds to the
 region indicated by dashed lines.}
\label{fig:bragg}
\end{figure}
As a specific example, we consider a Bragg stack as shown in
Fig.~\ref{fig:setup}b, consisting of alternate layers of
Poly(methyl methacrylate) (PMMA) and the chiral liquid. PMMA is
transparent to UV light down to approximately 300~nm, which is a
requirement for most applications involving biochemical molecules.
In Fig.~\ref{fig:bragg}a we show the band structure for a
Bragg stack of period $\Lambda=a_l+a_0$, where $a_l=0.8\Lambda$,
and $a_0=0.2\Lambda$. We use $n_l=1.33$ for the chiral liquid, and
a value of $n_0=1.7$ for the refractive index of PMMA.
\cite{Cardenas-Valencia2006} For all numerical results presented,
fully-vectorial eigenmodes of Maxwell's equations with periodic
boundary conditions were computed by preconditioned
conjugate-gradient minimization of the block Rayleigh quotient in
a planewave basis, using a freely available software package.
\cite{Johnson2001:mpb} We note that near the band-gap edges, the
first derivative of the dispersion relation approaches zero,
resulting in a strongly reduced group velocity
$v_g=\partial\omega/\partial k$. On the other hand,
{\highlight for the dielectric-like bands \cite{Mortensen2007a}}
the eigenmodes are strongly localized
in the high-index material
near the band-gap edges and thus $f\ll 1$. Nevertheless, as
seen in Fig.~\ref{fig:bragg}b, the enhancement
factor $\gamma$ still exceeds unity by an order of magnitude.
{\highlight This is discussed in more detail, and for 2D photonic
crystals such as those in Figs.~\ref{fig:setup}c and \ref{fig:setup}d,
in Refs.~\onlinecite{Mortensen2007a} and
\onlinecite{Mortensen2007}.}
In the inset of Fig.~\ref{fig:bragg}b we show the enhancement factor
as a function of wavelength in the UV region. The wavelength is
calculated for a structure with a period $\Lambda=250$~nm, which
is attainable using existing fabrication methods.
\cite{Christiansen2006} We note that for the
considered example the enhancement factor exceeds unity by an
order of magnitude for wavelengths within the UV range.

\begin{figure}
\begin{center}
\includegraphics[width=.95\linewidth, trim = 0 0 0 0, clip]
{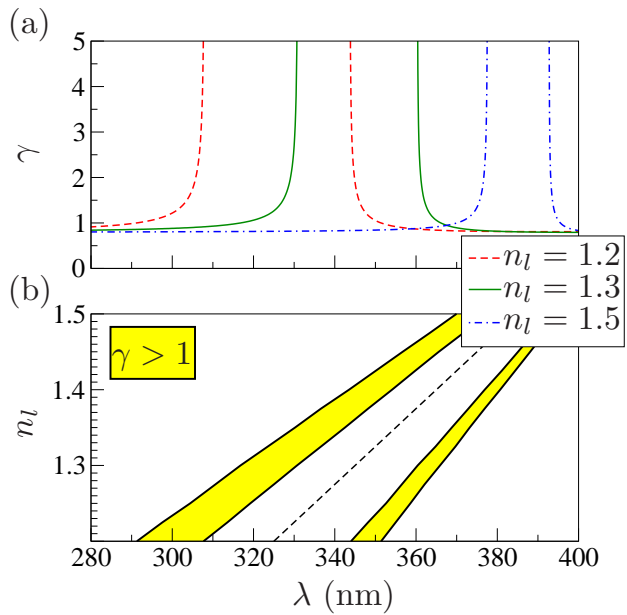}
\end{center}
 \caption{(Color online) (a) Enhancement factor for the Bragg
 stack of Fig.~\ref{fig:bragg} as a function of wavelength
 in the UV region, for three different values of the
 refractive index $n_l$ of the liquid. (b) Contour plot
 of the enhancement factor $\gamma(\lambda, n_l)$. The
 regions in which $\gamma$ exceeds unity are indicated. The
 dashed line indicates the wavelength corresponding to the middle
 of the band gap.}
\label{fig:gamma}
\end{figure}
In biochemical applications, the refractive index of the liquid
may vary depending on the specific buffer in use. It is thus
instructive to investigate the dependence of these results on the
refractive index of the liquid. Recently, Erickson
\emph{et al.} showed how the optical properties of planar photonic
crystal structures can tuned by varying the liquid refractive
index.\cite{Erickson2006} The effect is conveniently analyzed
within the framework of electromagnetic perturbation
theory\cite{Mortensen2007a} and as the refractive index is
increased, spectral features are generally redshifted. The
enhancement factor for the Bragg stack of Fig.~\ref{fig:bragg} is
shown in Fig.~\ref{fig:gamma}a, as a function of wavelength in the
UV region for three different values of the refractive index of
the liquid. As the index contrast is decreased, the photonic
band-gap edges are redshifted and the band gap size is decreased.
However, the enhancement factor still exceeds unity within the UV
region for all three values of $n_l$. In Fig.~\ref{fig:gamma}b we
show a contour plot of the enhancement factor $\gamma$
in a $\lambda$ versus $n_l$ diagram, indicating the regions in
which $\gamma>1$ as the refractive index of the liquid is
varied in the range from $n_l=1.2$ to $n_l=1.5$. As
the index contrast is decreased, the regions of $\gamma>1$ become
narrower, but even for a very low index contrast, the enhancement
factor still exceeds unity within the UV region. In the figure we
have also indicated the position of the center of the relevant
band gap, $\lambda_c$, as a function of the refractive index of
the liquid. In this particular case, the enhancement is due to the
second-lowest band gap, the center of which can be approximated
excellently by $\lambda_c\simeq a_ln_l+a_0n_0$
obtained by extrapolating the long wavelength limit dispersion
relation. The wavelengths for which the enhancement factor
exceeds unity thus grow approximately linearly with the refractive
indices of either material of the Bragg stack, as is also clearly
evident in Fig.~\ref{fig:gamma}b. One can thus imagine tuning the
dispersion by altering the refractive index of the buffer liquid
in which the chiral molecules are deposited. This would allow for
alignment of the position of the desired enhancement factor with
the relevant absorption wavelengths of the chiral molecules under
investigation. Alternatively, one may use the strong
frequency dependence of $\gamma$ in combination with a tunable
laser source to optimize the $\gamma(\omega)L$ product to match
the enantiomeric composition. {\highlight The strong frequency
dependence of $\gamma$ means that the value of $\gamma(\omega)L$
may be tuned to match a wide range of enantiomeric compositions,
while keeping the frequency within the absorbance peak
of the molecules. Consequently, the detection range of current
CD measurements may be significantly increased.}
Finally, while the UV regime is
typically the preferred natural regime for CD we envision that the
slow-light phenomena can be used to enhance the otherwise even
more weakly chiral response in other frequency regimes, thus
potentially paving the way for efficient CD at also longer
wavelengths such as the visible and near-infrared regimes.

In conclusion, we have shown how {\highlight dispersive structured media}
may be used as
a means of enhancing measurements of circular dichroism. Due to
slow-light phenomena, the optical path length required for the
largest signal-to-noise ratio of a CD measurement may be reduced
by more than an order of magnitude. Conversely, these ideas may be
used to enhance the sensitivity of CD measurements while
maintaining the same optical path length. By tuning the dispersion
relation, the proposed ideas may be used
to accommodate different types of molecules with in
principle no loss of sensitivity.
{\highlight While we have only presented results
for a 1D Bragg grating here, similar results apply to 2D photonic
crystals, for which slow-light phenomena also occur.
\cite{Mortensen2007a, Mortensen2007}}

\emph{Acknowledgments.} We thank Jacob Riis Folkenberg for
stimulating discussions. This work is financially supported by the
Danish Council for Strategic Research through the Strategic
Program for Young Researchers (Grant No. 2117-05-0037).


\begin{thebibliography}{10}

\bibitem{Chen2003}
L. Chen, Y.~J. Zhao, F. Gao, and M. Garland, Appl. Spectrosc. {\bf
57},  797
  (2003).

\bibitem{Janasek2006}
D. Janasek, J. Franzke, and A. Manz, Nature (London) {\bf 442},
374  (2006).

\bibitem{Psaltis:2006}
D. Psaltis, S.~R. Quake, and C.~H. Yang, Nature {\bf 442},  381
(2006).

\bibitem{Monat:2007a}
C. Monat, P. Domachuk, and B.~J. Eggleton, Nature Photonics {\bf
1},  106
  (2007).

\bibitem{Erickson2007}
D. Erickson, Microfluid. Nanofluid. {\bf 3},    (2007),
\emph{Special issue on
  "Optofluidics"}, doi: 10.1007/s10404-007-0226-8.

\bibitem{Verpoorte2003}
E. Verpoorte, Lab Chip {\bf 3},  42N  (2003).

\bibitem{Mortensen2007a}
N.~A. Mortensen, S. Xiao, and J. Pedersen, Microfluid. Nanofluid.
{\bf 3},
  (2007), doi: 10.1007/s10404-007-0203-2.

\bibitem{UV-CD}
This type of setup is typical of commercial UV-CD apparatus, such
as for
  example the Jasco CD-2095, from which the illustration is derived. See
  http://www.jascoinc.com.

\bibitem{JASCO}
See e.g. http://www.jascoinc.com.

\bibitem{Mortensen2007}
N.~A. Mortensen and S. Xiao, Appl. Phys. Lett. {\bf 90},  141108
(2007).

\bibitem{Cardenas-Valencia2006}
A.~M. Cardenas-Valencia, J. Dlutowski, D. Fries, and L.
Langebrake, Appl.
  Spectr. {\bf 60},  322  (2006).

\bibitem{Johnson2001:mpb}
S.~G. Johnson and J.~D. Joannopoulos, Opt. Express {\bf 8},  173
(2001).

\bibitem{Christiansen2006}
M.~B. Christiansen, M. Sch{\o}ler, S. Balslev, R.~B. Nielsen,
D.~H. Petersen,
  and A. Kristensen, J. Vac. Sci. Technol. B {\bf 24},  3252  (2006).

\bibitem{Erickson2006}
D. Erickson, T. Rockwood, T. Emery, A. Scherer, and D. Psaltis,
Optics Letters
  {\bf 31},  59  (2006).

\end{thebibliography}

\end{document}